# Proton-Pump Mechanism in Retinal Schiff Base: On the molecular structure of the M-state

Ayan Datta and Swapan K. Pati [*]

*Theoretical Sciences Unit and Chemistry and Physics of Materials Unit, Jawaharlal Nehru Center for Advanced Scientific Research, Jakkur P. O, Bangalore-560064, India.*

**Abstract**

Theoretical characterizations of the various intermediates in the proton pump cycle of the retinal Schiff base in the *Halobacterium salinarium* have been performed. Contrary to the general belief over the years that the most stable intermediate, the M-state, is a non-protonated cis-isomer, we find that the M-state is a polarized cis-isomer stabilized due to interactions of the dissociating proton with the π-electrons. The role of proton in the pump cycle is found to be profound leading to the stabilization or in certain cases destabilization of the intermediates. We propose the chemical structure of the M-state for the first time.

*Keywords:* Quantum chemical methods; Density functional calculations, Non-linear optical methods, excitation spectra calculations

Bacteriorhodopsin (bR), a transmembrane photosynthetic protein, is found in the purple membrane of *Halobacterium salinarium* [1,2]. On being excited by visible light, a proton-pump is triggered and the protein undergoes a cycle of events in which a proton is transferred from the cytoplasmic side to the extracellular side of the membrane, thereby creating a proton gradient across the membrane. Crystal structure reveals that this protein contains 7 transmembrane helices embedding the light absorbing chromophore, an all-trans retinal, covalently attached to Lys-216 by a protonated Schiff base linkage [3]. The important steps in the proton pump in bR involve an all-trans to 13-cis isomerization in the chromophore upon the excitation by light. In the next few steps, the proton is dissociated from this cis-isomer producing a stable intermediate, called M-state. After the release of the proton from the M-state, there is a conformational change in the chromophore involving a rotation along a single bond to produce the late M-state. This late M-state then accepts a proton from extracellular side and transforms to the ground state or the initial all-trans configuration, thereby completing the cycle.

However, a proper understanding of the processes in the membrane proteins remains elusive because of their difficult crystallization conditions [4]. Even the recent structural resolution of 1.5 Å is not sufficient to clearly show the complete atomistic picture of the various intermediates and particularly the position of the light atoms like hydrogen [5, 6]. Most of the theoretical efforts till now have taken as input the structures at various stages of the pump from these poor resolution X-ray patterns for further quantum mechanical or molecular dynamics calculations [7-9]. A full ab-initio calculation for the whole protein is not quite practical. As a result, there is no clear understanding of the essential factors that govern the proton pump cycle.

In this letter, we are able to show that a clear understanding of the whole pump cycle can be derived through a single universal parameter: the N...H distance. We find that the electrostatic interaction of the dissociating proton from this bond has a very important role in the generation of various intermediates in the proton-pump cycle and can cover the essence of the otherwise very complicated phenomenon in this biological membrane system. In particular, we explain the reason for the initial isomerization process of the ground state conformation in bR since generally it is known that the light induces a photoexcitation. The reason for the surprising fact that the cis-isomer is stable over the trans for the M-state, while in most cases a trans-isomer is generally found to be energetically stable, is explained. We sequentially explain the reasons behind each event as the Schiff bases pass through the pump cycle.

[*] Corresponding author. Tel: +91-80-22082829
E-mail: pati@jncasr.ac.in

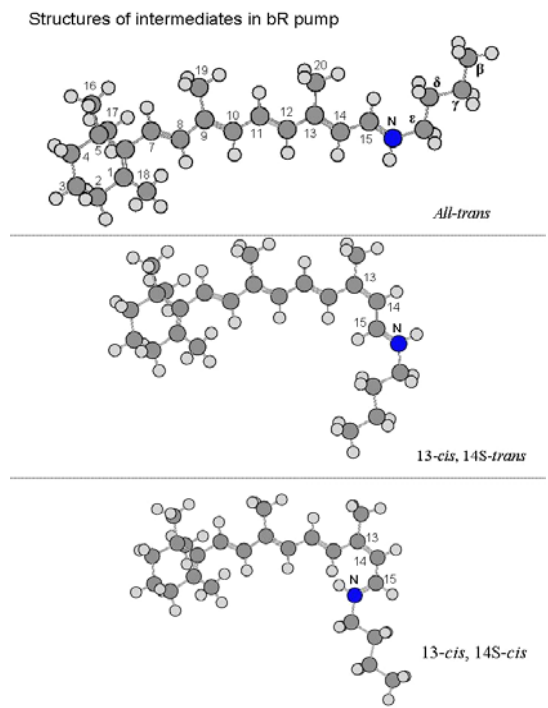

Fig. 1. Structures of the intermediates in bR. Note that the proton release from the N-atom in the 13-*cis*, 14S-*cis* occurs in the M-state.

We optimize the structures for the chromophores by substituting the Lys-216 residue by a butyl group. We have used the 6-31G(d, p) basis set and electron correlation was accounted by the Becke's three parameter hybrid method and the Lee-Yang-Parr correlation functional (B3LYP) [10]. Although our results are with butyl, we have verified that substituting butyl by lysine would give the same results with a very little quantitative difference. These geometries were used to compute the energies at various N...H distances.

In Fig. 1, the structures of various intermediates in bR cycle are shown. The first step in the process involves a all-trans to 13-cis, 14S-trans isomerization with a π-twist along the C(13)-C(14) bond which is carried out by the absorption of photon energy. We find that the lowest optically allowed state is almost 2.3eV above the all-trans ground state. This is calculated as the vertical absorption gap from the optimized ground state geometry of the all-trans isomer. A conformational isomerization (from optimized all-trans to π-twisted 13-cis, 14S-trans isomer) however requires only 0.17eV. In room temperature, although there will be a Boltzmann distribution of both the isomerized and photoexcited states, the number of bR molecules undergoing isomerization is quite large compared to the photoexcited state of the all-trans isomer. It is interesting to note that the energy required for this isomerization process in the protonated case is 0.17eV while it is 0.25eV for the deprotonated case. This is because to twist the double bond for the protonated case requires less energy as the formal +ve charge on the N drags the electrons and in the process the double bond attains a resonating single bond character. Starting from the 13-cis, 14S-trans isomer, we mimic the proton release process by elongating the N-H bond until the proton can be assumed to be free. At every N...H distance, we examine the change in energy due to the 'pulling' of proton from the N. The same profile is also examined for the all-trans isomer. We compare these two profiles in Fig 2(a). The all-trans isomer is always stable over the 13-cis, 14S-trans form, although the energy difference is quite small at all distances. Thus, the dissociating proton does not stabilize the 13-cis, 14S-trans state.

It has been suggested by Schulten *et al.* that after the twist along the C(13)-C(14) double bond, the bR under goes a 14S-trans to 14S-cis twist along a single bond to reach the 13-cis, 14S-cis (M-state) [11]. However, we find that the energy difference between these two structures is 1.3eV. Therefore, it appears surprising that such a step would occur, since it amounts to introducing steric repulsion in the system. To understand the reason behind the stability of the M-state and the actual conformation of the M-state, we compare the stability of this 13-cis, 14S-cis isomer with respect to the all-trans isomer for every N...H distance. The 'pulling' of proton for the 13-cis, 14S-cis case has been carried out by computing the lowest energy conformation at every N...H distance. We find that upto the N...H distance of 2 Å and after 3 Å, the most stable conformation is when the proton is out-of-plane of the molecule. In the intermediate distances, the stable conformation leads to proton being in the molecular plane [12]. Fig 2(b) describes the profiles for the all-trans and the minimum energy 13-cis, 14S-cis form at every N...H distance. It is quite clear from the figure that the all-trans isomer is lower in energy than the 13-cis, 14S-cis isomer for the associated as well for the dissociated N...H distances. But, for the case when the dissociating proton is in-plane with the 13-cis, 14S-cis isomer, some remarkable features appear. For the N...H distances between 2.4 Å and 3.2 Å, the 13-cis, 14S-cis isomer is more stable over the all-trans form, with optimum stability at d ~ 2.8 ° A. Therefore at such N...H distances, for the cis-conformation is stabilized. Since, this involves a dissociating proton, the stabilization must have an electrostatic origin. Thus, the distance region from d ~2.4 Å to d ~ 3.2 Å corresponds to an electrostatically significant distance.

Increase in the N-H distance in the Schiff base converts the $N^+...H$ having a formal +ve charge on the N to $N...H^+$, from where a proton is released eventually. Thus it is the role of the electric field created by the yet to be released proton that stabilizes the 13-cis, 14S-cis isomer. To confirm this hypothesis, we carry out similar calculations by substituting the Nitrogen in the Schiff base by Phosphorus (PH unit). The energy profiles for the all-trans and 13-cis, 14S-cis isomers of the P-Schiff base are shown in Fig. 2(c). The all-trans conformation is always stable over the 13-cis, 14S-cis isomer and nowhere in the energy surface is there a region of stable cis-conformer. This result fits naively into a very simple chemical intuition. P atom, having a low energy vacant d-orbital can accommodate pentavalency and a formal positive charge is thereby not developed. The process of dissociation of H occurs through a radical pathway unlike the case of the Schiff base where the dissociation is ionic with removal of $H^+$. Thus, one can claim conclusively that it is the $H^+$ on the verge of dissociation that stabilizes the 13-cis, 14S-cis form in bR.

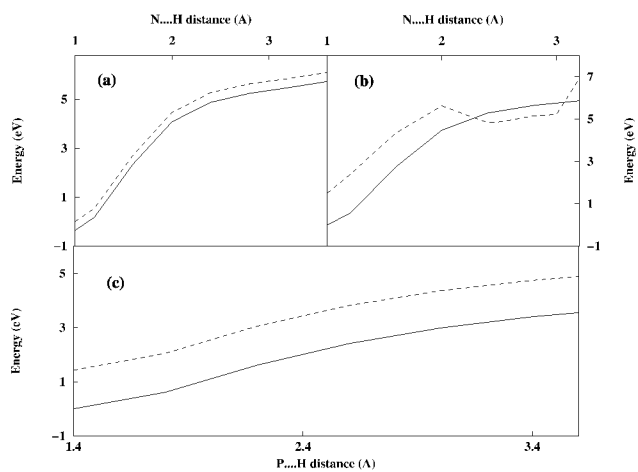

Fig. 2. Energy profiles as a function of proton distance for all-*trans* (solid line) and (a) 13-*cis*, 14S-*trans* (dashed line), (b) 13-*cis*, 14S-*cis* (dashed line). (c) all-*trans* (solid line) and 13-*cis*, 14S-*trans* (dashed line) isomers of P-ylide. Energies are scaled as E=E+23784.5 eV for (a) and (b) and as E=E+31601.6 eV for (c).

At the optimum N...H distance, d ~2.8 Å, the dissociating proton directs along the –C(13)=C(14)-C(15)=N planar moiety. For this particular distance, the proton forms the vertex of a 5 member ring geometry for the 13-cis, 14S-cis isomer. This structure is energetically stable since the geometry allows the proton to interact with $4\pi$ electrons. On the otherhand, for the all-trans form at similar distances, no such intermediate is formed as the proton is far away from the C(13)=C(14)-C(15)=N moiety. This also explains the reason for the specific trans to cis isomerization along the C(13)=C(14) bond for bR.

In Fig. 3, the LUMO wavefunction for the 13-cis, 14S-cis isomer is plotted at a N...H distance of 2.8 Å. The polarization effects are most evident in the LUMO wavefunction as it corresponds to the excited state structure with proton being pulled away. As can be seen from the plot, for the 13-cis, 14S-cis isomer, there is a ring like structure formation wherein the dissociating proton interacts with the π-electrons of the C(13)=C(14)-C(15)=N unit, thereby stabilizing the 13-cis, 14S-cis isomer.

The fact that the formation of such a cyclic intermediate can stabilize a cis-isomer is known for long in cyclo-addition chemistry (Diels-Alder Reactions). The stereochemistry of these reactions require the diene to be in a cis-conformation. But, since a trans diene is more stable over a cis form, the reaction rate increases for cases where one can lock the diene in a cis conformer eg. cyclopentadiene, furan [13].

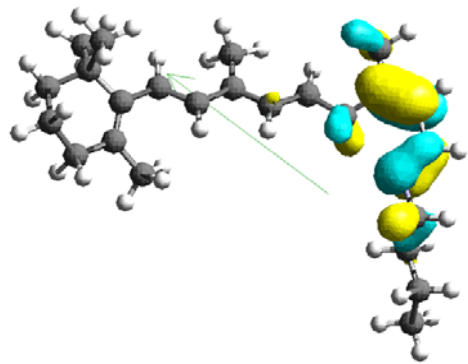

Fig. 3. LUMO wavefunction plot for the 13-cis, 14S-cis structure with N…H distance=2.8 Å. Note the localization of coefficients at the cis C(13)=C(14)-C(15)=N region. The light green arrow shows the direction of the ground state dipole moment.

The reaction rate is also significantly increased in the presence of Li + , K + or Lewis-Acids where the metal ions interacting with π-electrons can stabilize a cis diene [14,15].

As a further proof of our conjecture of stabilization of a cis isomer over the trans isomer through the ring formation, we perform an ab-initio calculation at B3LYP/6-311G++(d, p) level on a small model system: cis and trans 1,3-butadiene. The energy profiles for both the isomers are shown in Fig. 4. The proton is brought from one side towards π-electron region of the diene and taken in the opposite direction by varying the distance from the intermolecular axis of the diene (see Fig. 4). For the trans isomer, the profile is symmetric as in both the cases (front and back face), the interaction is isotropic. But, the cis isomer shows that in the front face between d ~2.0 Å to d ~ 3.0 Å, when the proton interacts with the π electrons of C=C-C=C, the cis form is stable over the trans isomer (see the inset). Thus, a proton forming a five-membered ring with a cis diene can stabilize a cis isomer over the trans isomer, a fact already seen for bR.

Now we discuss the next step after the M-state formation in bR where the proton is released. In order to investigate the stable geometries at the dissociated limit of the proton from the Schiff base, we minimize the geometries of the cist-conformations at large N...H distances by freezing that bond. We find that at such large N...H distances, the 13-cis, 14S-cis form isomerises back to the 13-cis, 14S-trans form. In this limit, the 13-cis, 14S-trans form is stable by 0.25 eV from the 13-cis, 14S-cis form. This observation suggests that at small N...H distances, the electrostatic interaction of the proton stabilizes the otherwise unstable 13-cis, 14S-cis form. As the proton starts to dissociate from the N, the M-state gets progressively unstable as the planarity of the 5 member ring is lost. This results in the back-isomerisation of the 13cis, 14S-cis form to the 13-cis, 14S-trans form. This conformation corresponds to the late M-state.

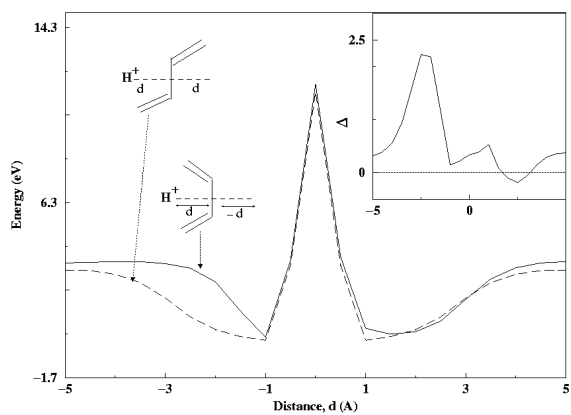

Fig. 4. Energy profiles for trans (dashed line) and cis (solid line) 1,3butadiene with respect to H+ distance. The dashed line across the molecule is the proton path. Energies are scaled as E=E+4247.1 eV. Inset shows delta=E(cis)-E(trans) vs the distance, d.

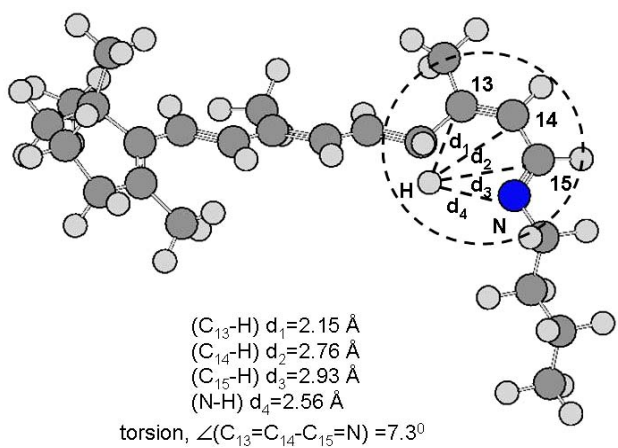

$(C_{13}-H) d_1 = 2.15$ Å
$(C_{14}-H) d_2 = 2.76$ Å
$(C_{15}-H) d_3 = 2.93$ Å
$(N-H) d_4 = 2.56$ Å
torsion, $\angle(C_{13}=C_{14}-C_{15}=N) = 7.3^0$

Fig. 5. Geometrical features for M-state in bR. The dotted circle shows the π-electron 'polarization pocket' due to interactions with H+.

Since, our calculations reveal that the M-state structure is polarized; we would like to suggest the possible ways to characterize the M-state. The ground state dipole moment is expected to be high for a polarized system. However, we find that not only the ground state dipole moment, the transition dipole moment (oscillator strength) of the excited state is also quite high for the M-state. We have calculated the linear and nonlinear polarizabilities (α and β) [16,17] of these systems using the established ZINDO-MRDCI-CV quantum chemical formalism [18]. For the M-state, cis-isomer with $d_{N...H}$ = 2.8 Å, the α and β values are very high: 348 × $10^{-24}$ esu and 585784 × $10^{-30}$ esu respectively for a photon frequency of 1064 nm corresponding to Nd-YAG Laser. Both α and β increase sharply with the increase in photon frequency. Compared to the deprotonated cis-isomer, the magnitude of α (β) for our suggested M-state is 12(777) times higher. Therefore, we strongly suggest that the measurement of the linear and nonlinear polarization should give a very precise idea of the nature of M-state. Infact, the M-state is being actively used for holography and in many other solid-state devices [19].

Finally, we suggest the chemical structure of the M-state by performing a geometry optimization for the 13-cis, 14S-cis isomer by varying the N...H distance. Indeed we find that the M-state corresponds to a local minima in energy with N...H distance=2.56 Å and an additional torsional twist of 7.30 in the 13-cis, 14S-cis conformation (Fig. 5).

To conclude, we have been able to identify the chemical structures of the intermediates in the proton-pump cycle of bR for the first time. In this work we considered only the Schiff-base portion. Particularly, we find that the M-state is not a neutral species but has a proton at an electrostatically significant distance which essentially stabilizes the M-state and accounts for its long lifetime. A more rigorous study including QM/MM method for the whole protein is being pursued and also the role of solvation/biological water is being investigated.

Acknowledgments: SKP thanks CSIR and DST, Govt. of India, for the research grants.